# Measuring photoexcited electron and hole dynamics in ZnTe and modeling excited state core-valence effects in transient XUV reflection spectroscopy

Hanzhe Liu,[1,*] Jonathan M. Michelsen,[1,*] Isabel M. Klein,[1] Scott K. Cushing[1,†]

[1]Division of Chemistry and Chemical Engineering, California Institute of Technology, Pasadena, CA 91125, USA.

[*]These authors contributed equally.

[†]Corresponding author. Email: scushing@caltech.edu

**Abstract**

Transient XUV spectroscopy is growing in popularity for the measurement of solar fuel and photovoltaic materials as it can separately measure electron and hole energies for multiple elements at once. However, interpretation of transient XUV measurements is complicated by changes in core-valence exciton and angular momentum effects after photoexcitation. Here, we report the photoexcited electron and hole dynamics for ZnTe, a promising material for $CO_2$ reduction, following 400 nm excitation. We apply a newly developed, ab-initio theoretical approach based on density functional theory and the Bethe-Salpeter equation to accurately predict the excited state change in the measured transient XUV spectra. Electrons excited to the conduction band are measured with a thermalization rate of $70 \pm 40$ fs. Holes are excited with an average excess energy of ~1 eV and thermalize in $1130 \pm 150$ fs. The theoretical approach also allows an estimated assignment of inter- and intra-valley relaxation pathways in k-space using the relative amplitudes of the core-valence excitons.

Photoelectrochemical reduction and oxidation reactions are driven by the transfer of a photoexcited electron or hole, respectively[1–3]. Knowing the initial excess energy of hot carriers, their thermalization times, and their recombination times is important for understanding and optimizing the desired reaction. It is therefore vital to measure photoexcited electron and hole dynamics separately[4–8]. Transient extreme ultraviolet (XUV) spectroscopy is a rising tool that can fill this role[4–8]. The core-valence XUV transition is sensitive to changes from photoexcited carriers and can also infer some structural dynamics. A picture of the coupled electronic and structural dynamics from the first few femtoseconds to the longer timescales of carrier recombination can be formed[7,9–13]. Multiple elemental X-ray absorption edges can then be used to measure electron or hole transport dynamics in the multilayer junctions required for modern photoelectrochemical cells[14,15].

The complex electronic-structural dynamics measured by transient XUV spectroscopy also bring new challenges for interpreting photoexcited spectra. The XUV probe photon excites a core-level electron to a valence state, leaving a positively charged core hole. This core hole acts as a perturbation to the final state such that the XUV absorption or reflection spectrum does not always mimic the unperturbed density of states[9,11,14,16,17]. Directly assigning hole and electron features on the simple basis of increases or decreases in absorption from the change in valence occupation is often impossible. New theoretical approaches that can incorporate the photoexcited changes in the predicted XUV transition are needed for accurate interpretation[16,18,19]. These methods are especially necessary when performing transient spectroscopy in a reflection geometry where changes in the real and imaginary parts of the refractive index must be considered[6,13].

In this study, we use transient XUV reflection spectroscopy to measure the ultrafast electron and hole dynamics in ZnTe, a material of growing interest for $CO_2$ reduction[20–23]. Transient XUV reflection spectra are measured at the Te $N_{4,5}$ edge around 40 eV following photoexcitation with a ~50 fs, 400 nm pulse. A Bethe-Salpeter equation approach is developed and used to theoretically predict the effects of photoexcited carrier distributions on the measured XUV spectrum[18]. The ab-initio approach allows for a rigorous reconstruction of the transient XUV reflection spectra without the need for toy models. The energy of the photoexcited electrons and holes versus time and the relaxation pathways of the photoexcited holes towards the Γ point are assigned. The hot electron relaxation rate to the conduction band minimum is measured to be 70 ± 40 fs, while the hot hole relaxation dissipates ~1 eV of extra energy in 1130 ± 150 fs. The results are relevant for the development of ZnTe photoelectrodes, particularly when it comes to nanoscale heterostructures and plasmonic approaches. Even more generally, we demonstrate an ab-initio approach to interpret transient XUV measurements without the need for system-specific approximate models[10,11], greatly expanding the technique's utility in complex solar energy materials and junctions.

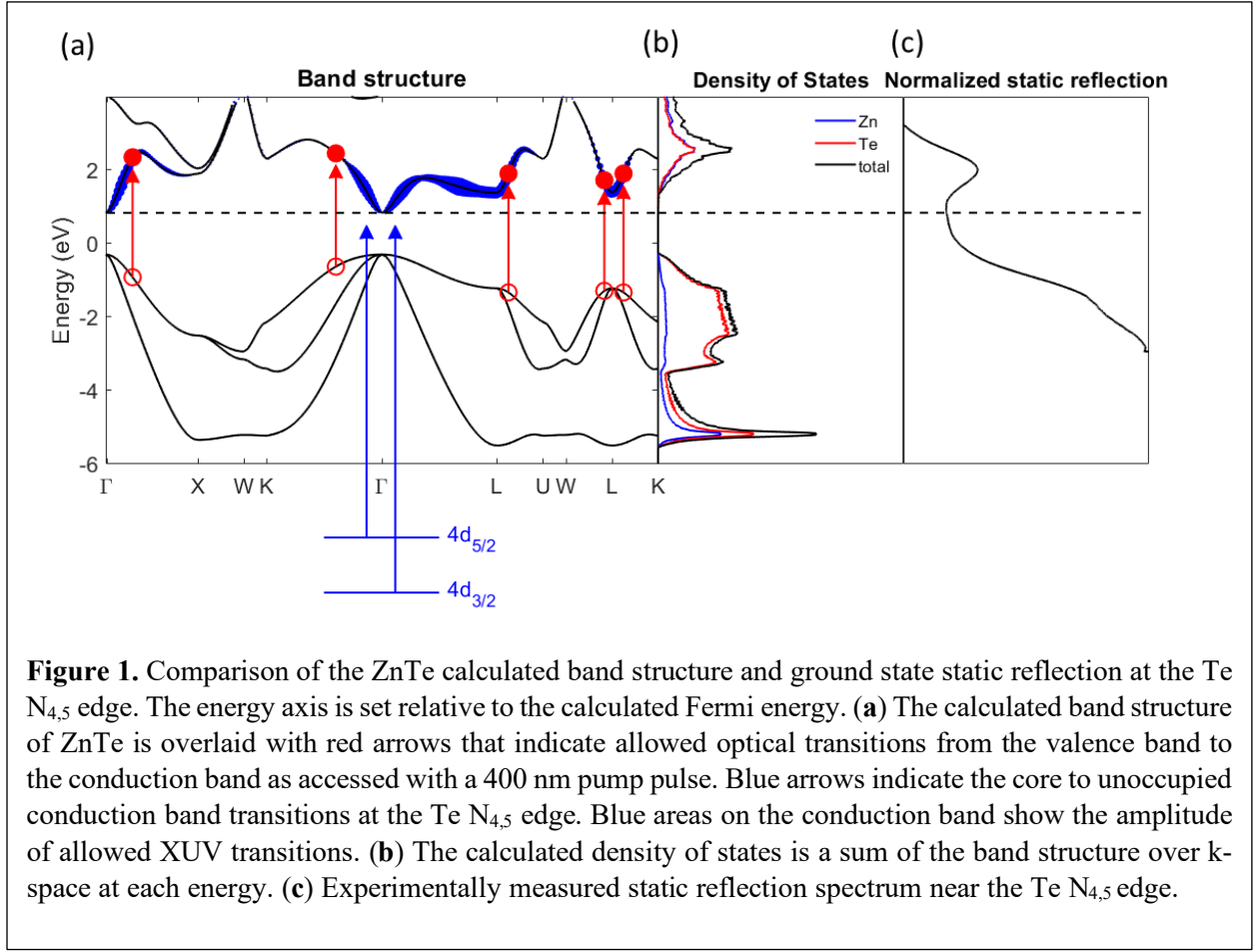

**Figure 1.** Comparison of the ZnTe calculated band structure and ground state static reflection at the Te $N_{4,5}$ edge. The energy axis is set relative to the calculated Fermi energy. (**a**) The calculated band structure of ZnTe is overlaid with red arrows that indicate allowed optical transitions from the valence band to the conduction band as accessed with a 400 nm pump pulse. Blue arrows indicate the core to unoccupied conduction band transitions at the Te $N_{4,5}$ edge. Blue areas on the conduction band show the amplitude of allowed XUV transitions. (**b**) The calculated density of states is a sum of the band structure over k-space at each energy. (**c**) Experimentally measured static reflection spectrum near the Te $N_{4,5}$ edge.

In this study, the sample is photoexcited by a ~50 fs, 400 nm frequency doubled output of a BBO crystal with p-polarization, pumped with an 800 nm, 1 kHz regeneratively amplified Ti-sapphire laser. The red arrows overlaid on the calculated band structure of ZnTe in Figure 1(a) indicate the allowed optical transition from the valence band to the conduction band accessed by the 400 nm pump pulse. The excitation flux is 1.74 mJ/cm$^2$, resulting in an initial photoexcited carrier density of 5.41 × 10$^{20}$ cm$^{-3}$. The photoexcited carrier dynamics are then probed with an XUV pulse produced by high-harmonic generation in argon with an s-polarized few-cycle white light pulse (~6 fs, 550 nm - 1000 nm, see Supporting Information). The generated XUV pulse covers an energy range between 30 eV to 70 eV, probing the Te $N_{4,5}$ edge as schematically indicated by blue arrows on Figure 1(a). The XUV probe is a sum over the dipole allowed transitions across k-space. The residual white light beam is removed with a 100 nm thick Al filter. The reflectivity measurement uses a 10-degree grazing incidence geometry. A typical XUV spectrum is shown in Supporting Information Section 2. The static sample reflectivity around the Te $N_{4,5}$ edge is shown in Figure 1(c). The reflectivity decreases around the conduction band minimum (around 40 eV from core states).

Photoexcited changes to the reflectivity are measured by varying delay times between the excitation and probe pulses using an optomechanical delay stage. The change in the transient reflectance after optical excitation is calculated by

$$\Delta OD = -\log_{10}\left(\frac{I_{pump\,on}}{I_{pump\,off}}\right) \quad (1).$$

The resulting experimental transient XUV reflection spectra is shown in Figure 2(a). In a transient absorption measurement, an increase in signal following photoexcitation could indicate increased transitions while a decrease in signal would represent the blocking of existing transitions. In the reflectivity spectrum, an increase in absorption will generically lead to a decrease in reflectivity, complicated by the contribution of both the real and imaginary components of the refractive index. From equation (1), the negative change in the color map (blue) in Figure 2(a) are a transient increase in reflectivity after photoexcitation, while the positive signal (red) is a transient decrease in reflectivity. The main transient features include an increase in reflectivity between 40 eV and 41 eV as well as between 41.5 eV and 42.5 eV. A decrease in reflectivity is measured starting at 38 eV and blue shifted to 39.5 eV approximately 1 ps after optical excitation. The initial increase in reflectivity below 40 eV blue shifts and disappears in time.

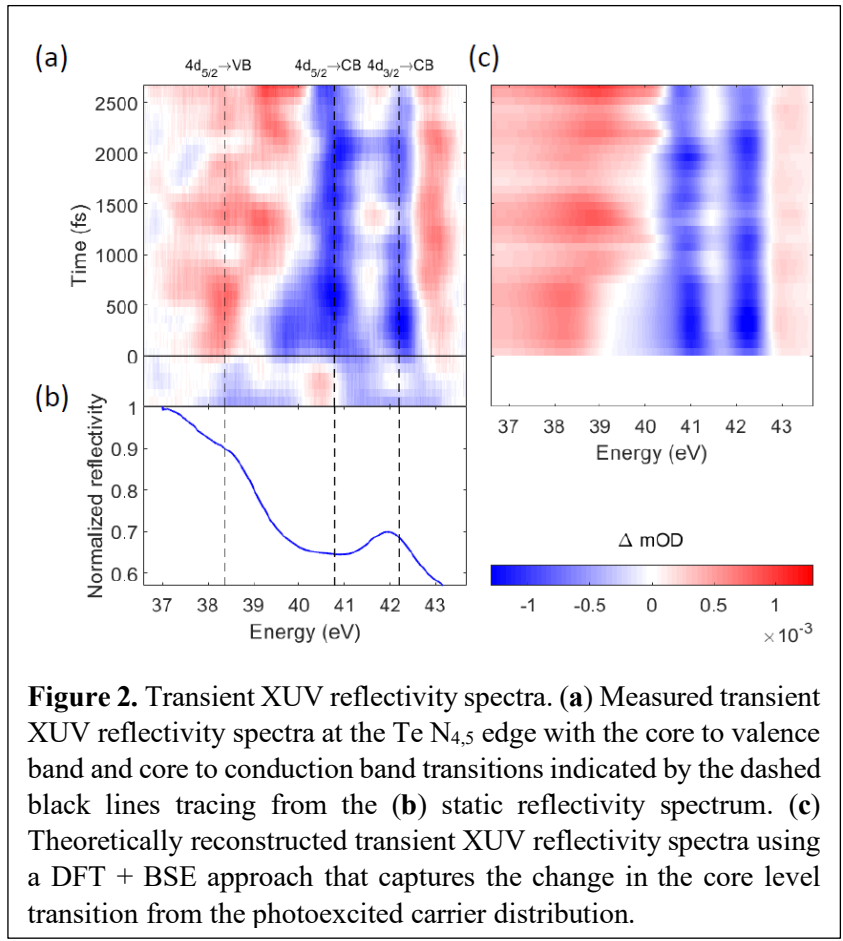

**Figure 2.** Transient XUV reflectivity spectra. (**a**) Measured transient XUV reflectivity spectra at the Te $N_{4,5}$ edge with the core to valence band and core to conduction band transitions indicated by the dashed black lines tracing from the (**b**) static reflectivity spectrum. (**c**) Theoretically reconstructed transient XUV reflectivity spectra using a DFT + BSE approach that captures the change in the core level transition from the photoexcited carrier distribution.

The optical excitation promotes electrons to the otherwise empty conduction bands, blocking possible XUV transitions in the transient spectrum. The creation of photoexcited holes allows new transitions to the valence band. Since a negative $\Delta$ mOD (blue) in Figure 2 corresponds to an increase in reflectivity, the spectrum qualitatively agrees with this approximation of how the transient spectrum should appear due to state blocking. The increase in reflectivity (blue color) in Figure 2(a) may correspond to the photoexcited electrons while the decreased reflectivity (red color) may correspond to photoexcited holes. However, this simplistic qualitative assignment would lead to the contradictory conclusion that the electrons with energies around 40 eV are increasing in energy over time instead of decreasing in energy.

To fully analyze the transient XUV spectra, state blocking and the associated changes in screening of the core-valence exciton must be considered, as well as any potential change in spin-orbit coupling and angular momentum effects[16,18]. These changes are calculated using an ab-initio combined theoretical approach based on density functional theory (DFT) and the Bethe-Salpeter equation (BSE)[18,19,24]. The existing OCEAN code[18,19] (Obtaining Core-level Excitations using ABINIT and the NIST BSE solver) is modified to accept excited state distributions as well as to determine how the ground state band structure relates to the observed XUV spectra and transient changes. In this procedure, the band structure and the ground state wavefunction are first calculated using DFT (Quantum Espresso)[24]. The BSE is solved to obtain core-valence exciton wavefunctions including spin-orbit coupling and Coulomb screening of the core-valence exciton. The calculated XUV absorption spectrum is related to the experimental reflection spectrum through the Kramers-Kronig relations. An optional GW step can be included to correct band gaps and changes thereof due to electron-hole renormalization effects in the excited state, but here is replaced by a simple 0.5 eV shift determined from the measured spectra. Specific details of the ground and excited state calculations are given in the Supporting Information Section 3.

To predict the excited state transient spectra, the occupations of the band structure is changed, and the BSE calculation is rerun. The optical excitation creates electron and hole pairs at positions in momentum space where the energy difference between the conduction band and valence band equals the excitation photon energy (red arrows, Figure 1(a)). The transient XUV spectra is calculated for a wide range of possible k-space occupations (560 possible occupations) and then the calculated spectra (Figure 2(c)) are used to back-extract electron and hole distributions. The experimental transient XUV spectrum is matched with a high level of accuracy using this approach. Inclusion of changes in screening to the core-valence exciton and the combination of real and imaginary parts of the spectrum (SI Section 4) successfully accounts for the spectral changes that would appear contradictory in the approximate approach described in the previous paragraph.

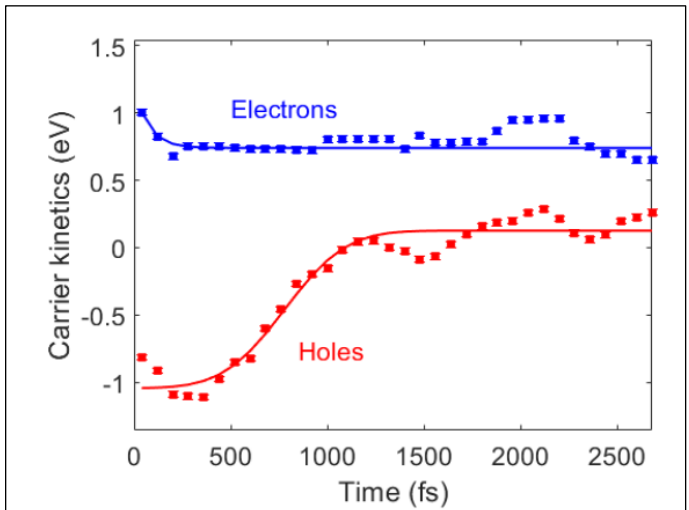

**Figure 3.** Extracted carrier kinetics from transient XUV reflectivity spectra using the Bethe-Salpeter Equation approach. The electron lineout (blue squares) has a relaxation rate of $70 \pm 40$ fs. The hole lineout (red squares) has a hole thermalization time of $1130 \pm 150$ fs from an initial energy of $1.2 \pm 0.1$ eV below the valence band maximum.

With an accurate prediction from the DFT + BSE theoretical framework, the electron and hole kinetics can be extracted from the experimental data and the kinetics fit to retrieve thermalization times (Figure 3). The electrons are excited with minimal excess energy and thermalize to the conduction band minimum in $70 \pm 40$ fs. Meanwhile, holes are excited with an average $1.2 \pm 0.1$ eV excess energy and thermalize in $1130 \pm 150$ fs. Note the reported energies are an average over the XUV transitions across k-space, analyzed in more detail in the following paragraph.

Knowing the k-space distribution of the core-valence exciton amplitudes in the ground and excited state allows for an estimate of excited state k-space occupations with time (Figure 4). To be clear, unlike angle-resolved photoemission spectroscopy (ARPES)[25], transient XUV spectroscopy does not measure the momentum distribution in k-space versus time for electrons and holes. Rather, the excited k-space occupations are back-estimated based on the predicted spectrum that matches the experimental data. For example, Figure 4 shows the extracted initial (t=0, red) and final (t=2.6 ps, blue) hole distributions in k-space. Immediately after photoexcitation, the measured holes are in a nonequilibrium distribution between the Γ and K points as well as in regions near the L point consistent with the allowed optical transitions (red circles and arrows). A significant hole density is not measured along the Γ to X direction although optical and XUV transitions are possible. At 2.6 ps after photoexcitation (blue) the holes have thermalized to the valence band maximum around the Γ point, noting that a transition exactly at the Γ point appears dipole disallowed for the XUV probe. The electrons (not shown) have minimal excess energy. Again, it should be emphasized that the k-space distribution is only probed where there is an XUV allowed transition so the hole assignments are only approximate.

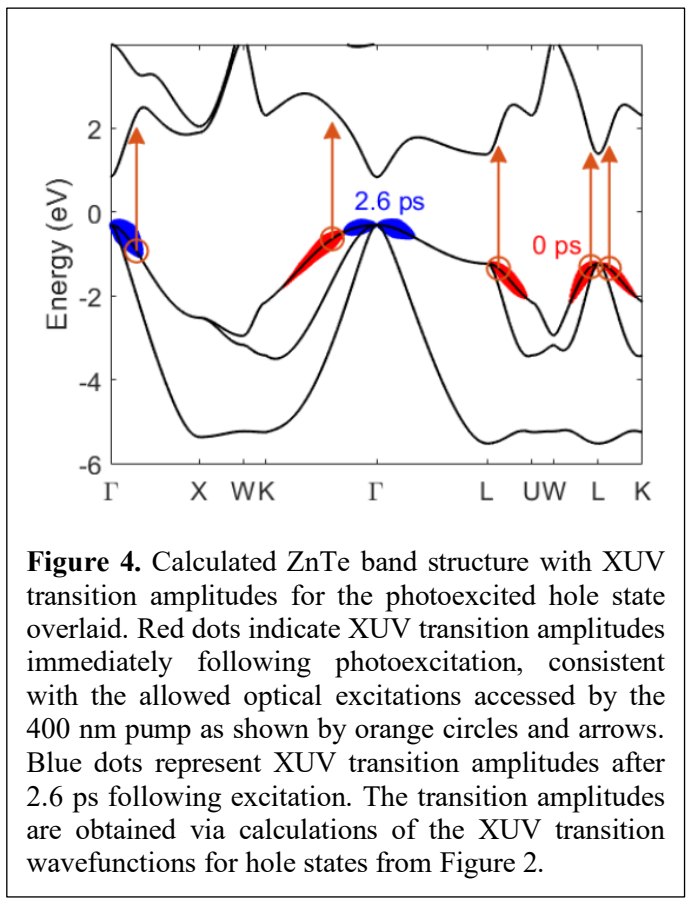

**Figure 4.** Calculated ZnTe band structure with XUV transition amplitudes for the photoexcited hole state overlaid. Red dots indicate XUV transition amplitudes immediately following photoexcitation, consistent with the allowed optical excitations accessed by the 400 nm pump as shown by orange circles and arrows. Blue dots represent XUV transition amplitudes after 2.6 ps following excitation. The transition amplitudes are obtained via calculations of the XUV transition wavefunctions for hole states from Figure 2.

In this investigation, we measure the transient XUV reflectivity of ZnTe at the Te $N_{4,5}$ edge. An ab-initio DFT-BSE framework incorporates the photoexcited effects in the core-level spectra, including angular momentum and core-valence excitons, allowing back-extraction of the electron and hole kinetics. The electron and hole dynamics are therefore extracted without the need for various excited state approximations or simplified models as used in past works[9–11]. For 400 nm excitation, electrons are measured to have minimal excess energy and a thermalization rate of 70 ± 40 fs. Holes are measured to have ~1 eV of extra energy and a thermalization time of 1130 ± 150 fs. The simultaneous measurement of electron and hole dynamics in ZnTe heralds future studies of more complex ZnTe heterostructures[20,21,23] and plasmonic catalysts for $CO_2$ reduction[26–28]. Just as critically, the experimental validation of the DFT-BSE theoretical framework for excited state predictions provides a path forward for interpreting and predicting the photoexcited changes in transient XUV experiments of complex materials and materials systems.